\documentclass[times,12pt,a4paper]{article}

\usepackage[a4paper,left=2.5cm,right=2.5cm,bottom=3cm,top=2.5cm]{geometry}

\usepackage[numbers, sort&compress]{natbib}

\usepackage[table]{xcolor}
\usepackage{amsmath}
\usepackage{amssymb}
\usepackage{amsthm}
\usepackage{graphicx}

\usepackage{fltpage}

\usepackage{listings}
\usepackage[colorinlistoftodos]{todonotes}
\usepackage{booktabs}

\usepackage{lscape}

\definecolor{lightgray}{gray}{0.9}

\usepackage{moreverb,url}

\usepackage{setspace}
\makeatletter
\let\@fnsymbol\@arabic
\makeatother

\begin{document}

\title{Optimal exact tests for multiple binary endpoints}

\author{Robin Ristl\footnote{Center for Medical Statistics, Informatics, and Intelligent Systems, Medical University of Vienna, Austria} , Dong Xi\footnote{Novartis Pharmaceuticals Corporation, East Hanover, NJ 07936, U.S.A.} , Ekkehard Glimm\footnote{Novartis Pharma AG, Basel, Switzerland}\textsuperscript{\ \ ,}\footnote{Otto-von-Guericke-University Magdeburg, Germany} , Martin Posch\footnotemark[1]}


%
%


\newcommand{\bt}{{\boldsymbol{t}}}
\newcommand{\bx}{{\boldsymbol{x}}}
\maketitle

\begin{abstract}
\noindent In confirmatory clinical trials with small sample sizes, hypothesis tests based on asymptotic distributions are often not valid and exact non-parametric procedures are applied instead. However, the latter are based on discrete test statistics and can become very conservative, even more so,  if adjustments for multiple testing as the Bonferroni correction are applied.
We propose improved exact multiple testing procedures for the setting where two parallel groups are compared in multiple binary endpoints. Based on the joint conditional distribution of test statistics of Fisher's exact tests, optimal rejection regions for intersection hypotheses tests are constructed. To efficiently search the large space of possible rejection regions, we propose an optimization algorithm based on  constrained optimization and integer linear programming. Depending on the optimization objective, the optimal test yields maximal power under a specific alternative, maximal exhaustion of the nominal type I error rate, or the largest possible rejection region controlling the type I error rate. 
Applying the closed testing principle, we construct optimized multiple testing procedures with strong familywise error rate control. Furthermore, we propose a greedy algorithm for nearly optimal tests, which is computationally more efficient.
We numerically compare the unconditional power of the optimized procedure with alternative approaches and illustrate the optimal tests with a clinical trial example in a rare disease.
\end{abstract}


%

\newpage
\section{Introduction}
In small confirmatory clinical trials, asymptotic hypothesis tests often do not provide strict control of the type I error rate. For many testing problems, exact tests based on conditional inference and permutation tests have been proposed. For such tests, the null distribution of the test statistic is typically discrete and the nominal significance level is not fully exhausted. Furthermore, in many settings more than one hypothesis is tested and a multiple testing procedure is applied to control the familywise type I error rate (FWER) \cite{EMAmultiplicity}. However, multiple testing procedures  that do not take into account the discreteness of the elementary exact tests, often become even more conservative. One remedy are randomized tests. They, however, are not accepted in practical applications such as clinical trials and will not be further considered here.

In this paper we construct optimized exact tests of intersection hypotheses and construct multiple testing procedures for elementary hypotheses based on the closed testing principle \cite{marcus1976}, focusing on multiple Fisher's exact tests for binary endpoints.

Several approaches to account for the discreteness of test statistics in multiple testing procedures have been proposed. For example, the Bonferroni test  can be improved, using the fact that for discrete tests a lower bound for the elementary p-values exists. Therefore, one can exclude hypotheses from testing (and multiplicity adjustment) for which this lower p-value bound  exceeds the nominal familywise significance level \cite{gart1979statistical}. This idea was refined by Tarone \cite{tarone1990modified}, who noticed that tests whose lower p-value threshold is larger than the corresponding adjusted level of the Bonferroni test, do not contribute to the type I error rate. While Tarone's test controls the FWER, it is not $\alpha$-consistent. A test is said to be $\alpha$-consistent, if the rejection of the null hypothesis at some significance level $\alpha$ implies that it can be rejected at all significance levels greater than $\alpha$.
Hommel and Krummenauer \cite{Hommel1998673} and Roth \cite{roth1999multiple} derived a more powerful and $\alpha$-consistent procedure by applying Tarone's procedure for all levels $\alpha' \leq \alpha$ and rejecting a null hypothesis if Tarone's procedure rejects for any $\alpha'\leq \alpha$.
Type I error rate control of all these procedures holds by the Bonferroni inequality. Hence, they rely on the marginal distributions of the test statistics only and do not take into account their joint distribution. 
Even though the latter is typically unknown, in the important case of between-group comparisons, the conditional joint distribution of the test statistics can be found by permutation of the group labels. The resulting distribution is conditional on the observed data.

Inference on the global null hypothesis of all marginal null hypotheses being true may be performed by combining the marginal test statistics or marginal p-values into a univariate statistic according to a pre-specified function. The null distribution of this statistic may be determined from the joint permutation distribution. See \cite{pesarin2012review} for a review on combination of dependent tests in a multivariate permutation setting.

The permutation approach is also used in the well known minP test \cite{westfall1993}, that is also applicable to discrete tests \cite{westfall1989p}. 
In the minP test, for each permutation of the group labels, the minimum across the p-values of the marginal hypothesis tests is calculated. The global null hypothesis is rejected at level $\alpha$ if the proportion of permutations with a minimum p-value less or equal the actually observed minimum p-values is less or equal $\alpha$. 
This results in rejection regions of a particular shape that can be described as the complement of a hypercube in the space of observable test statistics or p-values.
In the case of discrete marginal tests, the probability mass of these rejection regions is typically below $\alpha$.  Additional events defined on the joint distribution of test statistics (or p-values) may be added to such regions, still maintaining a probabiltiy mass below or equal $\alpha$. Rom proposed one particular extension, allowing for rejection if the minimum observed p-value is larger than the usual minP threshold, given that the remaining ordered p-values are below appropriate thresholds \cite{rom1992strengthening}. A combination of the approach underlying Tarone's test and the minP test was suggested in \cite{Gutman20072380}.

In this paper we generalize these tests and  consider general rejection regions for multivariate exact test statistics. The rejection regions may have arbitrary shapes that are only constrained by certain regularity conditions. We determine optimized rejection regions to either maximize exhaustion of the nominal type I error rate, the power under a specific alternative point hypothesis or simply the number of elements in the region. To efficiently search for the optimal rejection region we propose a numerical optimization algorithm.

The idea to consider general rejection regions for discrete tests has been used before. Paroush proposed a test of a single, simple null hypothesis versus a simple alternative based on a univariate discrete test statistic. The rejection region that maximizes the power under the alternative and controls the type I error rate under the null hypothesis can be found by linear integer programming \cite{paroush1969integer}. Gutman and Hochberg extended this approach to the multidimensional case \cite{Gutman20072380} and defined discrete multivariate rejection regions for the vector of marginal test statistics to test a global intersection null hypothesis. 
They derived rejection regions that optimally exhaust the nominal type I error rate and discuss also optimization of the power.
However, the application of the approach in a closed testing procedure showed low power to reject elementary hypotheses. The authors attributed this to the lack of consonance of the procedure. However, it may rather be due to the fact that the used algorithm does not guarantee test decisions which are monotone in the marginal test statistics, such that a rejection with a certain observed effect does not imply rejection when a more extreme effect is observed.
Because of the potential non-monotonicity of the resulting rejection regions, exhaustion of the local type I error rate may not translate into high power. 

In this paper we extend the approach in \cite{Gutman20072380} in several ways: (i) we  introduce a monotonicity constraint in the optimization framework to guarantee that the rejection regions are monotone in the marginal test statistics; (ii) we  show how the monotonicty constraint facilitates to efficiently solve the optimization problem numerically with a branch and bound algorithm; (iii) we consider optimization for additional objective functions, including the power 
 and the size of the rejection regions; (iv) as generalization of the approaches based on Tarone's method, we construct optimally weighted Bonferroni tests; (v) we propose greedy algorithms as a computationally less demanding alternative to full optimization.

The paper is organized as follows. In Section \ref{sec:optregions} we introduce the optimization framework for general, discrete, multivariate test statistics. We distinguish the case of a known joint null distribution and the case where only the marginal distributions are known. We construct optimized tests for intersection hypotheses and derive multiple testing procedures for elementary hypotheses based on the closed testing principle \cite{marcus1976}. 
In Section \ref{sec:optfisher}, we apply the optimization framework to Fisher's exact tests for the comparison of two parallel groups in multiple binary endpoints and in Section \ref{sec:numeric.example} we illustrate the procedure with a numeric example \cite{lago2002ibuprofen}. In Section \ref{sec:power}, the unconditional power of the optimal exact tests for multiple binary endpoints is compared to alternative procedures in a range of scenarios. We close with a discussion on the proposed procedures for multiple binary endpoints and give examples for the application of the optimization framework to other multiple testing problems.

\section{Optimal rejection regions for discrete tests} \label{sec:optregions}
In this section we develop a general framework to determine optimal multivariate rejection regions for exact tests of an intersection null hypothesis based on a vector of $k$ discrete marginal test statistics.  
In Section \ref{sec:joint.distr}, we study the case of a known joint distribution of the $k$ test statistics. We describe the construction of rejection regions for a global intersection null hypothesis that are optimal with respect to a given optimization criterion. We consider in particular three optimization criteria: (i) exhaustion of the nominal type I error rate, (ii) maximizing the number of elements in the multivariate rejection region and (iii) maximizing the power of the resulting test under a specific alternative. In Section \ref{sec:Bonf} we study the optimal selection of critical thresholds in weighted Bonferroni tests, which only requires knowledge of the exact marginal distributions of the test statistics. In both cases, the optimal solutions are found through methods of numeric optimization. As an alternative, we propose in Section \ref{sec:greedy} greedy algorithms that provide approximately optimal solutions. In Section \ref{sec:MTP} we discuss the construction of multiple testing procedures based on the application of locally optimal intersection hypothesis tests in a closed testing scheme.

\subsection{Optimal rejection regions for intersection hypotheses} 

\subsubsection{Optimal general rejection regions based on the joint distribution of the test statistics} \label{sec:joint.distr}

Consider hypothesis tests of $k$ elementary null hypotheses $H_i, i=1,\hdots ,k$ with discrete test statistics $\boldsymbol{T}=(T_1,\hdots ,T_k)$ taking values in a finite set $V=V_1\times \hdots \times V_k \subseteq \mathbb{N}^k$ such that larger values are in favor of the alternative. 
We assume that the marginal distribution under $H_i$ of each $T_i$ and the joint distribution of $(T_1,\hdots ,T_k)$ under $H_0=\cap_{i=1}^k H_i$ are known.

We aim to  construct optimal rejection regions $R \subseteq V$ for $\boldsymbol{T}$ to test the intersection hypothesis $H_0=\cap_{i=1}^{k}H_i$ versus the alternative that at least one $H_i$ is false at a pre-specified level $\alpha$. Optimization is performed over all valid rejection regions, defined as subsets $R \subseteq V$  that satisfy
\begin{enumerate}
	\item[$(i)$] $P_{H_0} (\boldsymbol{T} \in R) \leq \alpha$,
	\item[$(ii)$] If $(t_1,\hdots ,t_k ) \in R$ then $\{(s_1,\hdots ,s_k ) \in V: s_1 \geq t_1,\hdots ,s_k \geq t_k \} \subseteq R$,
\end{enumerate}
where $P_{H_0}$ is the probability under the intersection null hypothesis $H_0$. Condition (i) establishes type I error rate control and condition (ii) is a monotonicity condition that ensures that whenever the test rejects for test statistics taking the values $\boldsymbol{t}=(t_1,\hdots ,t_k)$ it will also reject if one (or several) values of the elementary test  statistics are increased. 
\newcommand{\rregions}{\mathcal{R}}
\newcommand{\argmax}{\textnormal{argmax}}
\newcommand{\argmin}{\textnormal{argmin}}

Let $f:\{R: R \subseteq V\} \to \mathbb{R}$ define an objective function that assigns a real number to every rejection region $R\subseteq V$. Let $\rregions$ denote the set of all rejection regions $R\subseteq V$ that satisfy conditions $(i)$ and $(ii)$.  
Then the optimal rejection regions with respect to $f$ are given by
\begin{equation} 
R_{f}\in \argmax_{R\in \mathcal{R}} f(R)\label{eq:optim}
\end{equation}
In the numerical examples we consider three objective functions: 
To obtain the test that best exhausts the nominal type I error rate we choose the objective function 
\begin{equation} \label{eq:alpha}
f_A(R)=P_{H_0} (\boldsymbol{T} \in R)
\end{equation}

 An alternative objective function is the number of elements in the rejection region
 \begin{equation} \label{eq:volume}
 f_C(R)=|R|
 \end{equation}
where $|\cdot|$ denotes the cardinality of a set.  
Optimizing the exhaustion of the nominal type I error rate or the number of elements does not necessarily translate to an optimal power. To maximize the power under a specific alternative hypothesis we consider the objective function
\begin{equation} \label{eq:power}
f_P(R)= P_{H_A} (\boldsymbol{T} \in R), 
\end{equation}
where $P_{H_A}$ denotes the joint distribution of $\boldsymbol{T}$ under a specified alternative $H_A$.

\ \\
\noindent {\bf An algorithm to determine optimal rejection regions based on the joint distribution of test statistics}

The optimization problem (\ref{eq:optim}) can be written as a binary integer program that can be solved with a branch and bound algorithm. The algorithm below determines an optimal solution if the objective function $f$ satisfies $f(R)\leq f(R')$ for all sets $R,R'$ such that $R\subseteq R'$. This is, e.g., the case for the objective functions considered above.
We index the (vector valued) elements of the set $V$ such that $V=\{\boldsymbol{t}_i,i=1,\ldots, m\}$, where $m$ denotes the cardinality of $V$. Then we can represent a rejection region $R\subseteq V$ as a binary vector $\boldsymbol{x}=(x_1,\hdots,x_m) \in \{0,1\}^m$, where $x_i=1$ if $\boldsymbol{t}_i\in R$ and $x_i=0$ otherwise. Therefore, the objective function $f$ can also be defined as function on $\{0,1\}^m$ and we use both definitions interchangeably.

To solve (\ref{eq:optim}) we apply a branch and bound algorithm \cite{horst2013global}. For the algorithm, denote current partial solution vectors $\boldsymbol{x}$, with $x_i\in\{-1,0,1\}$, where $x_i=-1$ indicates that the algorithm has not yet decided whether the corresponding point belongs to the optimal rejection region, $x_i=0$ denotes that the point does not belong to the optimal region, and $x_i=1$ that the point belongs to it. The nodes of the branch and bound algorithm are now given by a partial solution vector $\boldsymbol{x}$, together with a lower and upper bound of the value of the objective function $f$. Furthermore, 
we denote the current best lower bound for the value of the objective function $f$ by $L$. Then the optimization algorithm is given by:
\newcommand{\ii}{{\hat{i}}}
\begin{enumerate}
	\item Initialize the set $S$ containing a single node with the solution $\boldsymbol{x}=(-1,-1,\hdots ,-1)$, lower bound $f(\emptyset)$ and upper bound $f(V)$ and set the current best lower bound to $L=f(\emptyset)$.
	\item Of all nodes in $S$ with not fully determined solution, let $N$ denote the node with the largest lower bound and $\boldsymbol{x}$ its current partial solution. (If there are several such nodes, any can be chosen.) Furthermore, let $\ii$ denote the index of the first entry of $\boldsymbol{x}$ equal to -1.
	\item Remove $N$ from $S$ and add two modified copies of $N$ to $S$: in the first, set $x_\ii=0$, in the second set $x_\ii=1$. For each of the two new nodes:
    \begin{enumerate}
	\item Check if, to satisfy condition $(ii)$, the chosen value for $x_\ii$ determines the value of other entries of $\boldsymbol{x}$ that are currently equal to $-1$. If so, set the required values in $\boldsymbol{x}$.
	\item Let $\boldsymbol{x'}$ (resp.~$\boldsymbol{x''}$) denote copies of $\boldsymbol{x}$ where all components equal to $-1$ are set to 0 (resp. 1). Set the lower bound of the node to $f(\boldsymbol{x'})$ and the upper bound to $f(\boldsymbol{x''})$.
	\item If $\sum_{i:x_i=1} P_{H_0}\left(\bt_i\right)>\alpha$ remove the node from $S$.
	\end{enumerate}
    \item Update $L$ to the maximum of the lower bounds of all nodes in $S$.
	\item Remove all nodes from $S$ with an upper bound $< L$.
	\item Repeat Steps 2 to 5 until $S$ contains only nodes with fully determined solutions. These are optimal solutions.
\end{enumerate}

P-values for tests with an optimized rejection region $R$ can be defined as follows. Let $r=|R|$ and as a starting point for the iteration let $R_r=R$. We have to distinguish two cases: (a) If the observed test statistic $\boldsymbol{t}_{obs} \in R$, iterate $R_{s-1}=R_s \backslash \boldsymbol{t}_s$, where $\boldsymbol{t}_s=\argmax \{P_{H_0}(\boldsymbol{t}): \boldsymbol{t} \in R_s, R_s \backslash \boldsymbol{t} \mbox{ meets condition } (ii)\}$ (and $\backslash$ denotes 'without'). Stop if $\boldsymbol{t}_s=\boldsymbol{t}_{obs}$ and set the p-value to $P_{H_0}(R_s)$. 
(b) If the observed test statistic $\boldsymbol{t}_{obs} \notin R$, iterate $R_{s+1}=R_s \cup \boldsymbol{t}_s$, where $\boldsymbol{t}_s=\argmin \{P_{H_0}(\boldsymbol{t}):\boldsymbol{t} \in V \backslash R_s, R_s \cup \boldsymbol{t} \mbox{ meets condition } (ii)\}$. Stop if $\boldsymbol{t}_s=\boldsymbol{t}_{obs}$ and set the p-value to $P_{H_0}(R_{s+1})$.

\ \\
\noindent {\bf Some comments on the optimization algorithm}

(a) Note that Step 3(a) is not part of the standard branch and bound algorithm and has a double impact. It ensures that the solutions satisfy the monotonicity condition $(ii)$ and at the same time it simplifies the optimization problem by reducing the number of solutions that need to be considered. Computationally, this step can be implemented by computing an $m \times m$ look up matrix $D=(d_{ij})$, where $d_{ij}=1$ if $\bt_j \geq \bt_i$  (where $\bt_j \geq \bt_i$ iff $t_{j,l} \geq  t_{j,l},l=1,\ldots,m$) and 0 otherwise. Then, in Step 3(a), if $x_\ii=1$, $x_j$ is set to 1 for all indices $j=1,\hdots ,m$ where $d_{\ii j}=1$. Similarly, if $x_\ii=0$, $x_j$ is set to 0 for all $j=1,\hdots ,m$ where $d_{j\ii}=1$.  Note that the algorithm can be further improved by pre-processing steps in which the search space is reduced by excluding points from $V$ according to simple necessary conditions following from conditions $(i)$ and $(ii)$ (see Appendix A). An R implementation of the branch and bound algorithm, the pre-processing and further functions to caclulate the optimal exact tests we describe, is provided in the online supplement.

(b) Note that the constraints $(i)$ and $(ii)$ are linear functions of $\bx$.
If the objective function $f$ can also be written as a weighted sum $\sum_{i=1}^{m} w_ix_i$ with appropriate weights $w_i$, the optimization problem can be formulated alternatively as a linear program, similar as in \cite{paroush1969integer} and \cite{Gutman20072380} (see Appendix B).
Therefore, in principle standard LP solvers, as, for example, lpsolve \cite{lpsolve}, which can be accessed through R \cite{R2013,lpsolvepackage}, can be used to solve the optimization problem. However, when using lpsolve on different numeric examples for optimizing rejection regions, we occasionally observed numeric issues resulting in non-optimal solutions or extremely long run times. According to personal communication with the maintainers of lpsolve, these may result from the involved probability values ranging across several orders of magnitude, hence proper scaling in the underlying simplex algorithm may be difficult. For the numeric calculations presented in this paper, our own implementation of the branch and bound algorithm was used. 

(c) In general, the optimization problem can have more than one solution. This can occur, for example, if the joint null distribution (and the distribution under the alternative) is symmetric in the endpoints. To reduce the set of solutions, optimization criteria can be combined and applied in a lexicographical order.

(d) If the search space $V$ is large and many points in $V$ have very small probability mass there are a large number of close to optimal solutions resulting in long computation times. In Appendix C we show how an approximately optimal solution can be found at substantially reduced computational cost if points with very small probability under the null hypothesis are handled separately in the algorithm. 

(e) Due to step 3(a), the proposed branch and bound algorithm only searches across potential solutions satisfying condition $(ii)$. If the search space is a $k$-dimensional hypercube $V_1\times \hdots \times V_k$, with $V_i$ the unidimensional range of values for $T_i$, there are $\frac{(\sum_{i=1}^k |V_i|)!}{\prod_{i=1}^k |V_i|!}$ such solutions. This number gives an upper bound for the number of points the branch and bound algorithm visits. For comparison, the standard branch and bound algorithm visits up to $2^d$ with $d={\prod_{i=1}^k |V_i|}$ points.

\subsubsection{Optimal rejection regions based on the marginal distribution of test statistics} \label{sec:Bonf}
The above optimization relies on the joint null distribution of the test statistics. If this distribution is unknown,  Bonferroni-type optimal multiple tests for $H_i,i=1,\ldots, k$ based on the exact marginal distributions can be derived that control the FWER at level $\alpha$ in the strong sense. 
The Bonferroni test rejects $H_i, i=1,\ldots,k$ if $T_i \geq c_i$, where $c_i$ are critical boundaries such that $\sum_{i=1}^k S_i(c_i) \leq \alpha$, and $S_i(t)=P_{H_i}(T_i \geq t)$ denotes the probability that the test statistic for the $i$-th endpoint is equal or exceeds $t$ under the marginal null hypothesis $H_i$. 

While for the unweighted Bonferroni test the $c_i$ are chosen such that $S_i(c_i) \leq \alpha/k,i=1,\ldots,k$, for the weighted test, the level $\alpha$ can be allocated across the hypotheses more flexibly. 
The framework of weighted Bonferroni tests includes for example Tarone's test \cite{tarone1990modified}, in which for some hypotheses $c_i$ can be chosen such that $S_i(c_i)=0$ and not all hypotheses are tested.
It also includes a method proposed by Westfall and Troendle in which a common value as small as possible for all $c_i$ is used, i.e. $c_i=\min\{c: \sum_{i=1}^k S_i(c) \leq \alpha\}$ \cite{westfall2008multiple}.

In general,
the critical boundaries $c_i$ can be chosen to meet some optimization criterion over the set of marginal rejection regions. Let $V_i \subseteq \mathbb{N}$ denote the set of values $T_i$ can take. We assume $|V_i|<\infty$ for all $i=1,\hdots ,k$. Then the marginal rejection regions are given by $R_i=\{t \in V_i: t\geq c_i\}$ and are defined by a vector of critical boundaries $\boldsymbol{c} \in V=V_1 \times \hdots \times V_k$. Let $g_i:V_i\to \mathbb{R}$ denote marginal objective functions that depend on the marginal rejection regions only and define the overall objective function $g=\sum_{i=1}^k g_i$.  Then the critical values optimizing $g$ are given by
\begin{equation} \label{eq:opt.problem.Bonf}
\boldsymbol{c}: g(\boldsymbol{c}) \rightarrow \mbox{max, s.t. } \sum_{i=1}^k S_i(c_i)\leq \alpha\,.
\end{equation}
The size of the search space for this optimization problem is bounded by $\prod_{i=1}^k|V_i^\alpha|$ where, $V_i^\alpha\subseteq V_i$ denotes the set of critical values $c\in V_i$ such that $S_i(c)\leq \alpha$. Thus, the search space is much smaller than for the optimization based on the joint distributions discussed above and the solution can be found either by an exhaustive search  or by integer linear programming (see Appendix D).

Objective functions for the marginal tests corresponding to the criteria (\ref{eq:alpha}) and (\ref{eq:power}) are, for example, the expected number of rejected elementary hypotheses under $H_0$
\begin{equation} \label{eq:marg.alpha}
g_A(\boldsymbol{c}) = \sum_{i=1}^k S_i(c_i)
\end{equation}
or the expected number of rejections under marginal alternatives $H_i^{(1)}$
\begin{equation} \label{eq:marg.power}
g_P(\boldsymbol{c})=  \sum_{i=1}^k P_{H_i^{(1)}}(T_i \geq c_i)
\end{equation}
These are upper bounds on the type I error rate and power to reject the intersection hypothesis $H_0$.

\subsubsection{Greedy optimization algorithms} \label{sec:greedy}
The above algorithms find an optimal solution, but they can be computationally demanding. As an alternative, greedy algorithms can be used to obtain approximate solutions that satisfy $(i)$ and $(ii)$. To determine a rejection region based on the joint null distribution of $\boldsymbol{T}$, define an objective function $f$ as in Section \ref{sec:joint.distr}. Start with the empty set $R_0$. Choose an operator $\mbox{opt} \in \{\mbox{argmax},\mbox{argmin}\}$. In an iterative manner, define $R_{s+1}=R_s \cup \boldsymbol{t}_s$, where $\boldsymbol{t}_s= \mbox{opt} \{f(R_s \cup \boldsymbol{t}): \boldsymbol{t} \in V \backslash R_s, R_s \cup \boldsymbol{t} \mbox{ meets conditions } (i) \mbox{ and } (ii)\}$. If no such point can be found, stop and the rejection region $R$ is given by the current $R_s$. The choice of the operator opt as $\argmax$ or $\argmin$ depends on whether maximal or minimal increments of $f(R)$ are aimed at. In the numeric examples the algorithm is applied with $f(R) = P_{H_0}(T \in R)$ and the $\argmin$ operator, attempting to obtain a region with good exhaustion of the nominal level and a large number of elements. In contrast, $f(R) = P_{H_A}(T \in R)$ and the $\argmax$ operator could be used when the objective is to maximize the power under $H_A$. The greedy algorithm results in $\alpha$-consistent tests by construction.

Similarly, a Bonferroni-type test as in Section \ref{sec:Bonf} with close to optimal exhaustion of the nominal level can be found by a greedy algorithm. Here, in each iteration an element is added to the rejection region of that marginal test, for which the resulting increment in the objective function is smallest (largest) and the overall level $\alpha$ is still controlled.

\newcommand{\bR}{{\bar R}}

\subsection{Multiple testing procedures} \label{sec:MTP}
In the small sample setting, the rejection of intersection hypotheses can be an important trial objective because the power to reject specific elementary hypotheses may be insufficient. However, in many applications rejection of elementary hypotheses will be of interest and we extend the optimal tests for intersection hypotheses to multiple testing procedures for the elementary hypotheses  $H_i,i\in I=\{1,\hdots ,k\}$ that control the FWER in the strong sense.

To derive a multiple testing procedure we construct optimal local level $\alpha$ tests for all intersection null hypotheses $H_J=\cap_{i \in J}H_i, J \subseteq I$ and then apply the closed testing principle \cite{marcus1976} to test the elementary hypotheses. The closed test rejects an elementary null hypothesis $H_i$ if all intersection hypotheses $H_J$ with $J\subseteq I,i\in J$ are rejected by the respective local level $\alpha$ test. 

Multiplicity adjusted p-values for intersection or elementary hypotheses $H_J$ in a closed test are defined by $p^{*}(J)=\max\{p(J') : J' \subseteq J\}$, where $J \subseteq I$ is an index set and $p(J')$ is the local p-value for the intersection hypothesis $\cap_{i \in J'} H_i$.

Even if each of the intersection hypothesis tests satisfies an optimality criterion, this does not imply that some optimality property holds for the overall closed testing procedure (see e.g. \cite{henning2015closed}). In particular, Gutman and Hochberg \cite{Gutman20072380} noted that closed tests with locally optimal discrete rejection regions need not be consonant such that the rejection of an intersection hypothesis does not necessarily imply the rejection of at least one of the elementary hypotheses \cite{gabriel1969simultaneous}. They attributed the observed low power to reject elementary hypotheses observed for their procedure to the lack of consoncance. However, as we do not observe a similar drop in power (see the numeric results below), we conjecture the low power might be due to the lack of a monotonicity constraint like condition $(ii)$ in their procedure.

Still, for a non-consonant test the power to reject at least one elementary hypothesis is lower than the power to reject the global null hypothesis. The derivation of consonant optimized closed tests for general testing problems is complex due to the large number of intersection hypotheses that need to be considered. For the case of two hypotheses, though, a minor modification of the optimization algorithm to derive the optimal test for $H_1 \cap H_2$ is sufficient to ensure consonance: Let $R_i,i=1,2$ be the one-dimensional rejection region of the marginal test for $H_i$. Then $B=\{V_1 \backslash R_1 \times V_2 \backslash R_2\} \cap V$ is the set of points in the multivariate search space $V$ where no elementary hypothesis is rejected. For a consonant procedure, the search space for an optimal rejection region must therefore be restricted to $V \backslash B$.
Maximizing objective functions (\ref{eq:alpha}) or (\ref{eq:power}) over this restricted search space results in tests with maximal  exhaustion of the FWER, or maximal power to reject at least one elementary hypothesis, respectively.

For the weighted Bonferroni tests, consonance is achieved for the general case of $k$ hypotheses if, starting from the global intersection hypothesis test, the critical boundaries for each marginal test statistic are non-increasing \cite{goeman2010sequential}. Formally, for all $J' \subset J$ and $i=1,\hdots ,k$, $c_i^{(H_{J'})} \leq c_i^{(H_J)}$ needs to hold, where $c_i^{(H_J)}$ is the critical boundary for the test statistic of the $i$-th endpoint in the test for $H_J$. This additional constraint can be easily implemented when optimizing the critical boundaries for the marginal tests, simply by reducing the search space accordingly.
It does not affect the power of the global test. As a consequence, the power to reject at least one elementary hypothesis is equal to the power to reject the global intersection hypothesis. When the critical boundaries for the local tests are found by the greedy algorithm for Bonferroni tests, the closed testing procedure is consonant by construction of the greedy algorithm.

\section{Optimized Fisher's exact tests for multiple binary endpoints} \label{sec:optfisher}
\newcommand{\mt}{\tilde{m}}
\newcommand{\mmt}{\tilde{\boldsymbol{m}}}
\newcommand{\MM}{\boldsymbol{M}}

\newcommand{\mm}{\boldsymbol{m}}
\newcommand{\yy}{\boldsymbol{y}}
\newcommand{\YY}{\boldsymbol{Y}}
\newcommand{\zz}{\boldsymbol{z}}
\newcommand{\ZZ}{\boldsymbol{Z}}
\newcommand{\qq}{\boldsymbol{q}}

We will now apply the algorithms of Section \ref{sec:optregions} to construct optimal testing procedures for multiple binary endpoints, making use of the permutation joint distribution of the vector of multiple Fisher's exact test statistics.

Consider a treatment ($Trt$) and control ($Ctr$) group with $n_g$ subjects in group $g \in \{Trt,Ctr\}$. The observations on the subjects are assumed to be independent within and between the groups. 
In the case of a comparison of these groups with respect to a single binary endpoint, the observations are independently Bernoulli distributed with success probability $p_{Trt}$ in the treatment group and $p_{Ctr}$ in the control group.
The observed data can be aggregated in a $2 \times 2$ cross-table and Fisher's exact test provides condititional exact inference on the null hypothesis $H: p_{Trt} \leq p_{Ctr}$ \cite{fisher1935logic,Agresti_Buch_2002}.
One of the four entries in the $2 \times 2$ table, say the number of successes in the treatment group, is chosen as test statistic $T$. Conditional on the table margins, $T$ has a hypergeometric null distribution and large values of $T$ are in favor of the alternative $p_{Trt} > p_{Ctr}$. Making the inference conditional on the observed margins removes the influence of the unknown nuisance parameter (e.g. $p_{Ctr}$, depending on the parametrization) and allows for an exact test for $H$.
Under the point null hypothesis $p_{Trt}=p_{Ctr}$ the hypergeometric distribution of $T$ is equivalent to the permutation distribution of $T$ that results from all permutations of the group labels.

For the case of $k$ binary endpoints, consider the null hypotheses $H_i: p_{i,Ttr} \leq p_{i,Ctr}, i=1,\hdots,k$, where $p_{i,g}$ is the marginal success rate in the $i$-th endpoint in group $g$. We are interested in one sided alternatives $p_{i,Ttr} > p_{i,Ctr}$ to establish a higher success rate of the new treatment compared to control. 

Regarding the joint observations in one patient with respect to all $k$ endpoints, there are $d=2^k$ possible outcome categories. We can formally define these categories by a set of index vectors $\mathcal{S}=\{(s_1,\hdots ,s_k):s_i \in \{0,1\}, i=1,\hdots ,k\}$ such that $s_i=1$ if the particular patient had a success in endpoint $i$. E.g. with $k=2$ endpoints,  $\mathcal{S}=\{(1,1),(1,0),(0,1),(0,0)\}$, indicating a success in both endpoints, endpoint 1 only, endpoint 2 only or neither endpoint. 
To simplify notation, these $d$ different outcome categories will be indexed by $s=1,\hdots,d$ in the following equations if not indicated otherwise.

The observations on the $j$-th patient, $j=1,\hdots ,n_g$, in group $g \in \{Trt,Ctr\}$ can be written as vector $\YY_{g}^{(j)} \in \{\zz \in \{0,1\}^d: \sum_{s=1}^d z_s=1\}$. The single non-zero entry $Y_{g,s}^{(j)}=1$ indicates that the patient is in the $s$-th outcome category. Let $q_{g,s}=P(Y_{g,s}^{(j)}=1)$ be the probability for this event. The distribution of $\YY_{g}^{(j)}$ is characterized by the vector $\qq_g=(q_{g,1},\hdots,q_{g,d})$ with $0<q_{g,s}<1$ and $\sum_{s=1}^d q_{g,s}=1$.

The data resulting from this model can be aggregated without loss of information in a $d \times 2$ contingency table with columns $\YY_{Trt}=\sum_{j=1}^{n_{Trt}} \YY_{Trt}^{(j)}$ and $\YY_{Ctr}=\sum_{j=1}^{n_{Ctr}} \YY_{Ctr}^{(j)}$ (see Table \ref{tab:example} below for an example), and row margins $\MM=\YY_{Trt}+\YY_{Ctr}$. 

As for marginal Fisher's exact tests, define $T_i=\sum_{(s_1,\hdots ,s_k): s_i =1} \YY_{Trt,s_1\hdots s_k}$ as the number of subjects in the treatment group with a sucess in endpoint $i$, and let $\boldsymbol{T}=(T_1,\hdots,T_k)$. Thus, $\boldsymbol{T}$ is a linear function $h$ of $\YY_{Trt}$. The distribution of $\boldsymbol{T}$ conditional on $\MM$ will be used for exact inference about $H_0 = \cap_{i=1}^k H_i$. 
This distribution is a function of the conditional distribution of $\boldsymbol{Y}_{Trt}$ given $\MM=\mmt$
\begin{equation} \label{eq:distrT}
P(\boldsymbol{T}=\boldsymbol{t}|\MM=\mmt)=\sum_{\boldsymbol{y}_{Trt} \in W:h(\boldsymbol{y}_{Trt})=\boldsymbol{t}} P(\boldsymbol{Y}_{Trt}=\boldsymbol{y}_{Trt}|\MM=\mmt)
\end{equation}
where $W=\{\yy \in \mathbb{N}^{d}: \sum_{s=1}^d y_s= n_{Trt} \mbox{ and } \exists \zz \in \mathbb{N}^{d}:  \yy+\zz=\mmt\}$ is the set values of $\yy_{Trt}$ that are possible given the table margins.
The conditional distribution of $\YY_{Trt}$ given $\MM=\mmt$ is a multivariate (non-central) hypergeometric distribution (see Appendix E)

\begin{equation} \label{eq:conddistr}
P(\YY_{Trt}=\yy_{Trt}|\MM=\mmt) = \frac{1}{N} \prod_{s=1}^d \frac{1}{y_{Trt,s}!(\mt_s-y_{Trt,s})!}\left( \frac{q_{Trt,s}}{q_{Ctr,s}} \right)^{y_{Trt,s}}
\end{equation}
with  normalizing constant $N=\sum_{\yy_{Trt} \in W}  \prod_{s=1}^d \frac{1}{y_{Trt,s}!(m_s-y_{Trt,s})!}\left( \frac{q_{Trt,s}}{q_{Ctr,s}} \right)^{y_{Trt,s}}$.

When $q_{Trt,s}=q_{Ctr,s}$ for all $s=1,\hdots,d$, (\ref{eq:conddistr}) is equivalent to the permutation distribution of $\YY_{Trt}$ given $\mmt$  that results from performing all possible permutations of the group labels and (\ref{eq:distrT}) is equivalent to the corresponding permutation distribution of $\boldsymbol{T}$. 
Similar to the minP approach \cite{westfall1993,westfall1989p}, we use this permutation distribution of $\boldsymbol{T}$ as null distribution under $H_0$. Optimal multivariate rejection regions are then determined by applying the optimization procedures of Section \ref{sec:optregions} over the search space $V=h(W)$.

Analogously, for a test of $H_J=\cap_{i\in J} H_i$, $J\subset \{1,\hdots,k\}$ the joint permutation distribution of $(T_i: i \in J), J\subset \{1,\hdots,k\}$ and the respective search space and optimal rejection regions are determined using only the data on endpoints $i \in J$.

When testing the intersection hypotheses $H_J$ by tests based on these permutation distributions, the closed testing procedure controls the FWER under the following additional exchangeability assumption, which is similar to the marginals-determine-the-joint condition given in \cite{xu2007applying}.

\ \\
\noindent $(iii)$ {\bf Assumption.} \textit{Let $K \subseteq \{1,\hdots,k\}$ be the index set of all true null hypotheses. The joint distribution of the observations on endpoints $i \in K$ is assumed to be identical in both treatment groups.} \\

\noindent FWER control follows, because under $(iii)$ the conditional null distribution of $(T_i: i \in K)$ given the respective table margins for the data on the endpoints $i \in K$, $\mmt_K$, is its permutation distribution. So the type I error rate of the permutation test for $H_K$ is controlled conditional on $\mmt_K$, and since this holds for any realization of $\mmt_K$ it also holds unconditionally. Type I error rate control of the test for $H_K$ is sufficient for FWER control of the closed test.
See \cite{westfall2008multiple,calian2008partitioning,klingenberg2009testing} for further discussion on testing multiple hypotheses using permutation tests and involved assumptions 
and \cite{mehta1994exact} for a general treatment of the permutation principle applied to contingency tables.

Calculating optimal Bonferroni-type tests, as described in Section \ref{sec:Bonf}, for multiple Fisher's exact tests is straight forward using the known marginal hypergeometric distribution of the individual test statistics $T_i, i=1,\hdots k$. Assumption $(iii)$ is not required for FWER control with the Bonferroni-type tests.
Note that, for both procedures, only the observed table margins $\mmt$ are required to compute the conditional null distribution.

\section{A clinical trial example} \label{sec:numeric.example}
For illustration, consider a trial to show superiority of ibuprofen compared to indomethacin in the treatment of patent ductus arteriosus in preterm infants, similar to the study described in \cite{lago2002ibuprofen}. The primary endpoint in this study was ductal closure and a major secondary endpoint was low urine output. For our example we will consider both binary endpoints as primary. As study outcome consider the observed frequencies given in Table \ref{tab:example}, that for either treatment group entail the same observed marginal success rates as in the original study (no information on the joint distribution is reported in \cite{lago2002ibuprofen}).

Assume the aim of this study is to show superiority of the new treatment compared to control in at least one of the two endpoints, at a familywise significance level of 2.5\%. The elementary null hypotheses are $H_{urine}: p_{urine,Trt} = p_{urine,Ctr}$ and $H_{duct.}: p_{duct.,Trt} = p_{duct.,Ctr}$, and the global intersection null hypothesis is $H_0=H_{urine} \cap H_{duct.}$. We assume exchangeability under $H_0$, according to $(iii)$.

\begin{table}[htbp]
  \caption{\bf Exemplary data for observed frequencies on two binary endpoints.}
  \centering
    \begin{tabular}{lccccc}
      & Treatment   & Control \\
     \midrule
	Success in both endpoints & 80    & 57\\
	Success in urine output only & 13  & 12 \\
	Success in ductal closure only & 1   & 10  \\
	Success in neither endpoint & 0     & 2 & \\
    \bottomrule
    \end{tabular}
  \label{tab:example}
\end{table}

To test the global intersecion hypothesis $H_0$ at level $\alpha=0.025$, we consider the test statistics from the marginal Fisher's exact tests for each endpoint. For this vector of test statistics, we calculate the permutation distribution and find rejection regions with optimal exhaustion of the nominal level, maximal number of elements and optimal power under an assumed alternative of independent endpoints with true success rates of $p_{urine,Trt}=p_{duct.,Trt}=0.9$ and $p_{urine,Ctr}=p_{duct.,Trt}=0.75$ (following the assumptions made for sample size planning in the original study). 
Analogous regions were caclulated under the additional constraint of providing a consonant closed test. In addition, the multivariate rejection region resulting from the greedy algorithm with the argmin operator was found. In the results Table \ref{tab:props} the respective tests are referred to as (consonant) optimal alpha, (consonant) optimal area, (consonant) optimal power and greedy algorithm.

Further, critical boundaries for optimal consonant Bonferroni-type tests with objective functions (\ref{eq:marg.alpha}) (Bonferroni optimal alpha) and (\ref{eq:marg.power}) (Bonferroni optimal power) were calculated, as well as boundaries resulting from the Bonferroni greedy algorithm. For comparison, the unweighted Bonferroni test, the Hommel-Krummenauer variant of Tarone's test (HKT) and the minP test were included. For the minP test, the minimum p-value across the marginal Fisher's exact tests was used as test statistic. The null distribution of this statistic was derived from the joint permutation distribution of the marginal test statistics, matching the permutation approach described in \cite{westfall1989p}.

\begin{figure}[h]
    \begin{center}
      \includegraphics[scale=0.35]{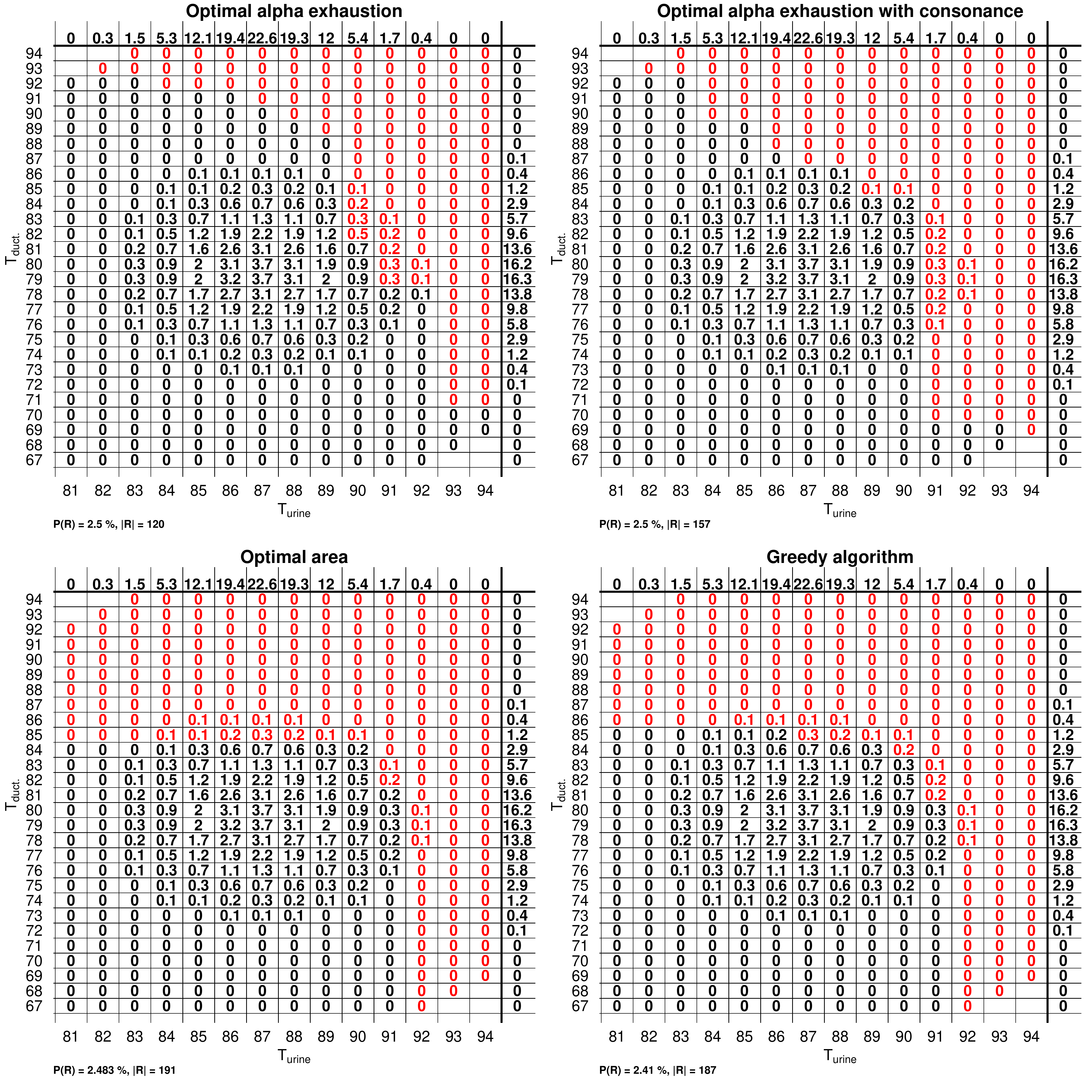}
    \end{center}
\caption{Rejection regions in the example of Section \ref{sec:numeric.example} with optimal alpha exhaustion, optimal number of elements and the region resulting from the greedy algorithm. The figure shows the conditional joint distribution of the Fisher's exact test statistics, $T_{urine}$ and $T_{duct.}$, for the two endpoints under the global null hypothesis and assuming exchangeability. Probabilities are given in percent and rounded to 0.1\%. Cells with entries 0 have a small positive probability, empty cells have probability 0. The upper and the right margins show the marginal distributions of $T_{urine}$ and $T_{duct.}$. The rejection regions are coloured in red. The probability mass of the rejection region ($P(R)$) and the number of elements in the region ($|R|$) are displayed below each graphic. The nominal significance level is 2.5\% for all tests.}
\label{fig:RegionsH0}
\end{figure}

\begin{figure}[h]
    \begin{center}
      \includegraphics[scale=0.35]{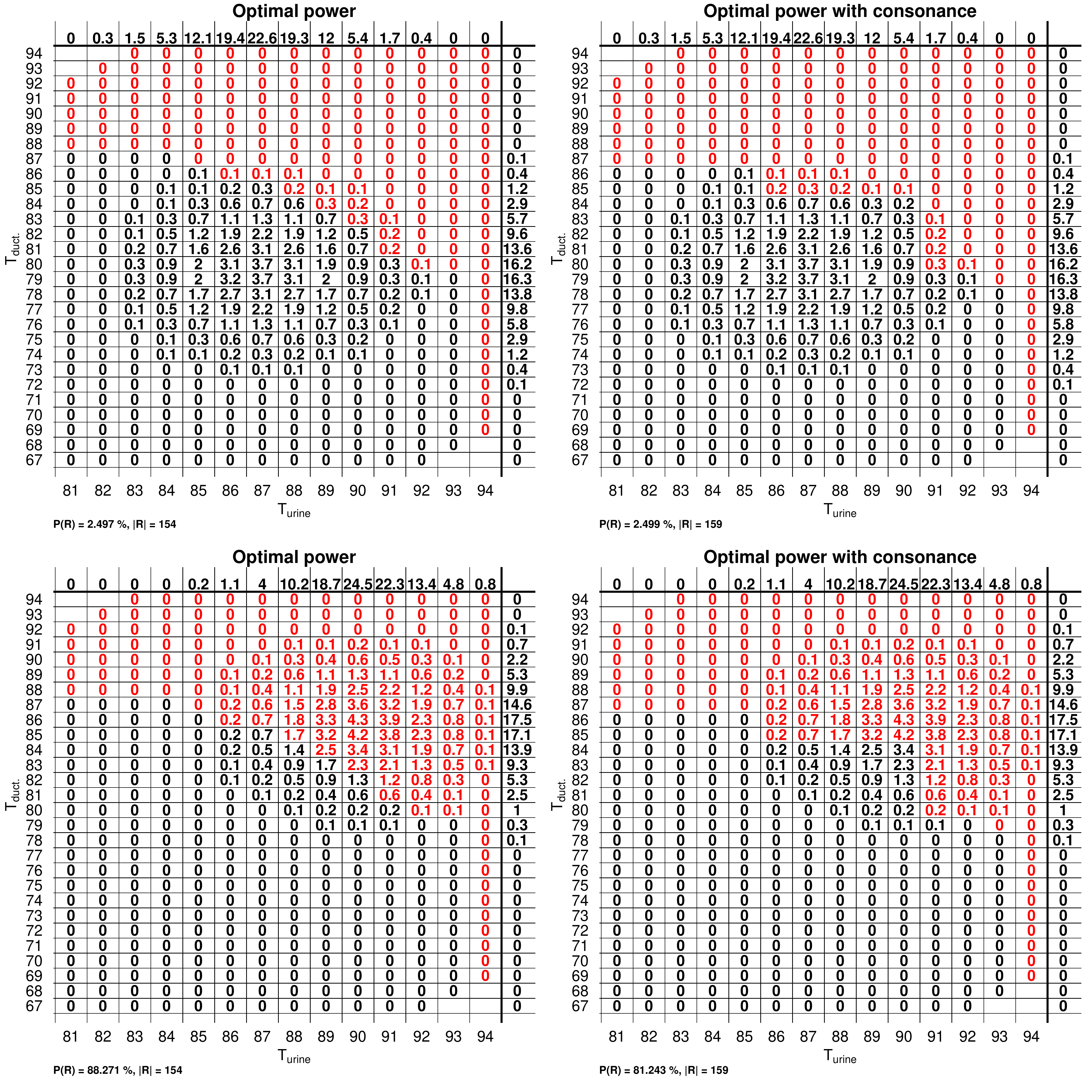}
    \end{center}
\caption{Rejection regions in the example of Section \ref{sec:numeric.example} with optimal power under the assumed alternative. The top row shows the conditional joint distribution of the Fisher's exact test statistics, $T_{urine}$ and $T_{duct.}$, under the null hypothesis, the bottom row shows the distribution under the assumed alternative. See also the legend of Figure \ref{fig:RegionsH0}.}
\label{fig:RegionsPower}
\end{figure}

For the included tests and conditional on the row margins of Table \ref {tab:example}, Table \ref{tab:props} shows the actual type I error rate under $H_0$, the power under the assumed alternative and the number of elements the rejection regions contain. For the Bonferroni-type tests and for the minP test, the critical boundaries are included in the table. The rejection regions for the proposed optimal tests based on the joint distribution are visualized in Figures \ref{fig:RegionsH0} and \ref{fig:RegionsPower}. Here, the tests that maximize alpha exhaustion or power are not consonant and Figures \ref{fig:RegionsH0} and \ref{fig:RegionsPower} include the respective rejection regions, when the additional constraint of consonance is imposed. 
In this example conservatism is greatly reduced and the conditional power increased when using optimal tests as compared to a basic Bonferroni test.

The marginal Fisher's exact tests reject at local level 2.5\% if $T_{urine} \geq 91$ and $T_{duct.} \geq 85$. The observed values for the test statistics in the example are $t_{urine}=93$ and $t_{duct.}=81$. The point $(93,81)$ is contained in all considered rejection regions (see Figures \ref{fig:RegionsH0} and \ref{fig:RegionsPower}), and so all examined tests reject $H_0$. Following the application of the closed testing principle all tests also reject $H_{urine}$, concluding that the proportion of patients with low urine output is lower under the new medication.

The p-value for $H_0$, calculated according to the suggestion in Section \ref{sec:joint.distr} after rejection of $H_0$ at the level $\alpha=0.025$, is approximately 0.0002 for the greedy algorithm test and the tests optimizing alpha exhaustion and area of the rejection region, with and without the consonance constraint. The marginal one-sided p-values using Fisher's exact test are $p_{urine}=0.0005$ and $p_{duct.}=0.3361$, and as both are larger than 0.0002 the multiplicity adjusted p-values take the same values, respectively. For the test optimizing the power under the specified assumption on the alternative, the p-value for the global test is 0.0006 without the consonance constraint and 0.0017 with the consonance constraint. When using these approaches, the multiplicity adjusted p-value for low urine output is equivalent to the p-value for the global test, the adjusted p-value for ductal closure again is 0.3361.

The results on the example data serve as illustration and apply only conditional on the specific observed margins.
A study of the unconditional properties of the proposed tests is given in the next section.

\clearpage
\begin{landscape}
\begin{table}[htbp]
  \centering
  \caption{{\bf Characteristics of different rejection regions for testing the global null hypothesis $H_0=H_{urine} \cap H_{duct.}$ in the example data set.} The table shows the type I error rate (Level)
 and power under the assumed alternative in percent, calculated conditionally on the observed total success numbers. The other columns show the number of points included in rejection regions ($|R|$) and 
the critical boundaries ($c_{urine}, c_{duct.}$) for the tests with rectangular-type rejection region that reject $H_0$  if $T_{urine} \geq c_{urine}$ or $T_{duct.} \geq c_{duct.}$. $|V|,\ |V^{(1)}|$ and $|V^{(2)}|$ are the number of elements in the original search space, initially and after the first and second pre-processing step, respectively (see Appendix A). The number of iterations that were required in the branch and bound algorithm to identify an optimal solution is given in the last column. The nominal type I error rate for all tests is 2.5\%. Tests that result in identical rejection regions in the example are reported in a single row.}
    \begin{tabular}{lccccccccc}
    \toprule
         Test  & Level & Power & $|R|$  & $c_{urine}$   & $c_{duct.}$  & $|V|$    & $|V^{(1)}|$   & $|V^{(2)}|$   & Iterations \\
    \midrule
    Bonferroni/HKT   & 0.98  & 60.3  & 177   & 92    & 86    &       &       &       &  \\
    Bonferroni optimal alpha & 2.27  & 61.3  & 186   & 91    & 87    &       &       &       &  \\
    Bonferroni optimal power/greedy/minP & 2.17  & 74.1  & 188   & 92    & 85    &       &       &       &  \\
    Optimal alpha & 2.50  & 66.8  & 120   &       &       & 386   & 212   & 159   & 357591 \\
    Optimal area & 2.48  & 80.6  & 191   &       &       & 386   & 212   & 159   & 5084 \\
    Optimal power & 2.50  & 88.3  & 154   &       &       & 386   & 212   & 159   & 60747 \\
    Consonant optimal alpha & 2.50  & 75.9  & 157   &       &       & 386   & 206   & 123   & 45317 \\
    Consonant optimal area & 2.48  & 80.6  & 191   &       &       & 386   & 206   & 123   & 1160 \\
    Consonant optimal power & 2.50  & 81.2  & 159   &       &       & 386   & 206   & 123   & 13014 \\
    Greedy algorithm & 2.41  & 84.3  & 187   &       &       &       &       &       &  \\
    \bottomrule
    \end{tabular}
  \label{tab:props}
\end{table}
\clearpage
\end{landscape}

\clearpage
\section{Unconditional power of the optimal procedures} \label{sec:power}
The unconditional power of different closed testing procedures based on either of the proposed optimal intersection hypothesis tests is studied for the setting of $k=2$ and $k=3$ binary endpoints. These intersection hypothesis tests are the optimally weighted consonant Bonferroni tests with objective function (\ref{eq:marg.alpha}) (labelled as Bonferroni optimal alpha) or objective function (\ref{eq:marg.power}) (Bonferroni optimal power), the Bonferroni greedy algorithm test with the argmin operator (Bonferroni greedy algorithm), optimal tests using the joint permutation distribution with the objective functions (\ref{eq:alpha}) (Optimal alpha), (\ref{eq:volume}) (Optimal area) and (\ref{eq:power}) (Optimal power) and the greedy algorithm test for joint distributions, using the argmin operator (Greedy algorithm). For the case of two endpoints, the corresponding optimal joint distribution-based tests with the additional constraint of consonance (Cons. opt. alpha, Cons. opt. area., Cons. opt power) are also studied.
These procedures are compared to closed testing procedures based on testing the local intersection hypotheses via the Bonferroni test, the Hommel and Krummenauer improvement of Tarone's test (HKT) or the minP test as described in Section \ref{sec:numeric.example}. In all cases, one-sided Fisher's exact tests are applied for the local elementary hypothesis
tests and the vector of the elementary test statistics was used as multivariate test statistic
in the intersection hypothesis tests.

The between-groups differences in these scenarios are parametrized by the marginal success rates $p_{i,Trt},i=1,\hdots,k$ in the treatment group and $p_{i,Ctr},i=1,\hdots,k$ in the control group. 
Further, a common product-moment correlation $\rho$ between the binary observations within each subject is assumed. Per-group sample sizes are $n_{Trt}=n_{Ctr}=n \in \{5,10,15,20\}$ for two endpoints and $n=10$ for three endpoints.

Table \ref{tab:scen2and3EP} shows the settings for all considered scenarios.
$p_{i,Trt}$ and $p_{i,Ctr}$ were chosen such that for endpoints where the alternative holds $(p_{i,Trt}+p_{i,Ctr})/2=0.5$ and such that the power to reject the hypothesis at an unadjusted level of 2.5\% using a single Fisher's exact test is approximately 0.6 (two endpoints) or 0.41 (three endpoints). 
Furthermore, one scenario with three endpoints with $p_{1,Trt}=0.8,\ p_{2,Trt}=0.7,\ p_{3,Trt}=0.6$ and $p_{i,Ctr}=0.2$ for all $i=1,2,3$, was considered. There, the corresponding local power values of Fisher's exact tests are 0.64, 0.43 and 0.25.

For the tests that directly aim to maximize the power, assumptions on the alternative need to be specified. For each scenario with two endpoints the optimal power tests are calculated under three different assumptions matching the three overall scenarios of an effect in one endpoint, an effect in both endpoints with $\rho=0$ and an effect in both endpoints with $\rho=0.5$. 
In this way, the true alternative is always included, and in addition the characteristics of the tests under assumptions that deviate from the truth can be assessed. For three endpoints the power was optimized under the true alternative.

The unconditional power was calculated numerically for the scenarios with two endpoints and by simulation for the scenarios with three endpoints. See the supplemental material for technical details.

\begin{table}[htbp]
  \centering
  \caption{Simulation scenarios for two and three binary endpoints.}
\begin{tabular}{cccccccccc}
\toprule
\#endpoints    &   n     & $p_{1,Trt}$ & $p_{2,Trt}$ & $p_{3,Trt}$ & $p_{1,Ctr}$ & $p_{2,Ctr}$ & $p_{3,Ctr}$ & $\rho$   & $\alpha$ \\
\midrule
2     & 5     & 0.865 & 0.135 &       & 0.135 & 0.135 &       & 0     & 0.025 \\
      &       & 0.865 & 0.865 &       & 0.135 & 0.135 &       & 0     & 0.025 \\
      &       & 0.865 & 0.865 &       & 0.135 & 0.135 &       & 0.5   & 0.025 \\
      & 10    & 0.792 & 0.208 &       & 0.208 & 0.208 &       & 0     & 0.025 \\
      &       & 0.792 & 0.792 &       & 0.208 & 0.208 &       & 0     & 0.025 \\
      &       & 0.792 & 0.792 &       & 0.208 & 0.208 &       & 0.5   & 0.025 \\
      & 15    & 0.265 & 0.265 &       & 0.265 & 0.265 &       & 0     & 0.025 \\
      &       & 0.265 & 0.265 &       & 0.265 & 0.265 &       & 0.5   & 0.025 \\
      &       & 0.735 & 0.735 &       & 0.265 & 0.265 &       & 0     & 0.025 \\
      &       & 0.735 & 0.735 &       & 0.265 & 0.265 &       & 0     & 0.025 \\
      &       & 0.735 & 0.735 &       & 0.265 & 0.265 &       & 0.5   & 0.025 \\
      &       & 0.297 & 0.297 &       & 0.297 & 0.297 &       & 0     & 0.05 \\
      &       & 0.297 & 0.297 &       & 0.297 & 0.297 &       & 0.5   & 0.05 \\
      &       & 0.703 & 0.297 &       & 0.297 & 0.297 &       & 0     & 0.05 \\
      &       & 0.703 & 0.703 &       & 0.297 & 0.297 &       & 0     & 0.05 \\
      &       & 0.703 & 0.703 &       & 0.297 & 0.297 &       & 0.5   & 0.05 \\
      & 20    & 0.701 & 0.299 &       & 0.299 & 0.299 &       & 0     & 0.025 \\
      &       & 0.701 & 0.297 &       & 0.299 & 0.299 &       & 0     & 0.025 \\
      &       & 0.701 & 0.701 &       & 0.299 & 0.299 &       & 0.5   & 0.025 \\
3     & 10    & 0.254 & 0.254 & 0.254 & 0.254 & 0.254 & 0.254 & 0     & 0.025 \\
      &       & 0.254 & 0.254 & 0.254 & 0.254 & 0.254 & 0.254 & 0.5   & 0.025 \\
      &       & 0.746 & 0.254 & 0.254 & 0.254 & 0.254 & 0.254 & 0     & 0.025 \\
      &       & 0.746 & 0.746 & 0.746 & 0.254 & 0.254 & 0.254 & 0     & 0.025 \\
      &       & 0.746 & 0.746 & 0.746 & 0.254 & 0.254 & 0.254 & 0.5   & 0.025 \\
      &       & 0.8   & 0.7   & 0.6   & 0.2   & 0.2   & 0.2   & 0.5   & 0.025 \\
\bottomrule
\end{tabular}
\label{tab:scen2and3EP}
\end{table}

\subsection{Numerical results}
The results for a selected scenario with a treatment effect in two uncorrelated endpoints and a sample size of $n=15$ per group are shown in Table \ref{tab:nEP2.n15.alpha0.025.corr0.effall.text}. All results on scenarios with two endpoints are tabulated in  the supplemental tables S1 to S19. 
The results for the scenario with three correlated endpoints with unequal effect sizes are shown in Table \ref{tab:power3EP.text}.
The results for the remaining scenarios with three endpoints are covered in the supplemental tables S20 to S25. 

Overall, the unweighted Bonferroni procedure had the lowest power in the considered scenarios. The Hommel and Krummenauer improvement of Tarone's test (HKT) differs from the Bonferroni test only for constellations where some tests cannot become significant at certain Bonferroni-adjusted levels $\leq \alpha$. This happens frequently only for very small sample sizes, therefore  the HKT test was substantially more powerful than the Bonferroni test only in scenarios with the very small sample size of $n=5$. For larger sample sizes the differences were small.

Conservatism was reduced and the power was notably increased by an order of 10 percentage points when the local boundaries in weighted Bonferroni tests were chosen according to one of the proposed optimization criteria or using the greedy algorithm. When the treatment effects in both endpoints were equal, there was almost no difference in the performance of these optimization approaches.
In the case with an effect in one endpoint only, however, optimizing power according to the objective function (\ref{eq:marg.power}) under the assumption of the true effects provided some additional advantage.

The minP test had very similar power values as the weighted Bonferroni tests. This is of particular interest, as the Bonferroni tests do not require Assumption $(iii)$ while the minP test does require the assumption to allow for unambiguous interpretation of tests on marginal effects.

The optimal tests based on the joint distribution provided a substantial improvement over HKT, the Bonferroni-type tests and over the minP test in the order of another 10 percentage points. By definition, optimizing the power under the true alternative results in the largest power, which can serve as a benchmark for the other tests.
When there was an effect in only one endpoint, optimizing exhaustion of the nominal level or optimizing power under different alternative hypotheses in some scenarios resulted in power similar to that of the Bonferroni test, though.
In contrast, maximizing the number of points in the rejection region gave more robust results, with power values above those of the weighted Bonferroni tests.

For the optimal tests using the joint distribution, enforcing consonance in the scenarios with two endpoints did not lead to a notable improvement of the power to reject at least one elementary hypothesis, but at the same time decreased the power to reject the global intersection null hypothesis. 

The tests obtained through the greedy algorithms, both for the Bonferroni approach and the joint distribution rejection region, performed surprisingly well. In most scenarios these tests had power similar to or above that of the respective other tests based on marginal or joint distributions.

Differences in the power characteristics between the testing procedures were mostly observed for the test of the global intersection hypothesis. For the elementary hypotheses, all tests other than Bonferroni and HKT showed similar power values, with a few exceptions in scenarios assuming an effect in just one endpoint. The probability to reject all elementary hypotheses simultaneously was almost identical for most of the studied tests. This is another consequence of the discreteness of the elementary tests, by which the set of values for the multivariate test statistic, that lead to the local rejection of all two or three elementary null hypotheses simultaneously, is often entirely contained in the optimal rejection regions. For an illustration see the numeric example of Section \ref{sec:numeric.example}. There, the set $\{(t_{urine},t_{duct.}): t_{urine} \geq 91, t_{duct.} \geq 85\}$ is contained in all optimal rejection regions. Even the simple Bonferroni rejection region almost completely contained this set, missing only the single point $(T_{urine}=91, T_{duct.}= 85)$.

For all tests other than unweighted Bonferroni and HKT, the power to reject a specific elementary hypothesis was typically very close to the local power that a single Fisher's exact test would have under the chosen marginal success rates. The power to reject at least one elementary hypothesis was even larger, with the exception of scenarios with an effect in only one endpoint. This observation implies that carefully accounting for the discreteness of the tests allows one to greatly reduce the cost of multiple testing. Thus, in the studied scenarios, the mulitplicity adjustment when testing two or three hypotheses does not reduce the power compared to the test of a single null hypothesis.

\begin{landscape}
\begin{table}[htbp]
\centering
\caption{Power with two uncorrelated binary endpoints with a treatment effect in both endpoints. The scenario for the table is $n=15$, $\alpha=0.025$, $p_{1,Trt}=0.735, p_{2,Trt}=0.735, p_{1,Ctr}=0.265, p_{2,Ctr}=0.265, \rho=0$.
The table shows the probabilities in percent that the closed testing procedure rejects the global intersection hypothesis $H_1 \cap H_2$, at least one elementary null hypothesis ($H_1$ or $H_2$), both elementary null hypothesis ($H_1$ and $H_2$), or particularly $H_1$, or $H_2$. The branch and bound algorithm was used with a maximal number of 20000 iterations.
To quantify the number of required iterations, the median (q50), the 90 \% quantile (q90) and the maximum of the number of iterations is shown in the last three columns. A value of 0 means that the optimization was finished by the pre-processing. 
For tests optimizing the power, the assumed alternative is indicated in brackets next to the test label. There, ``all'' and ``EP 1'' refer to an assumed effect identical to the true effect in a scenario with effect in both endpoints or one endpoint, respectively. $\rho$ there indicates the assumed correlation between the endpoints. 
}
\begin{tabular}{lcccccccccc}
    \toprule
    \multicolumn{1}{c}{Test} & $H_1 \cap H_2$ & $H_1$ or $H_2$ & $H_1$ and $H_2$ & $H_1$ & $H_2$ & q50 & q90 & Max \\
    \midrule
Bonferroni & 72.3 & 72.3 & 34.8 & 53.6 & 53.6 &   &  &  \\
HKT & 72.3 & 72.3 & 34.8 & 53.6 & 53.6 &  &    &  \\
Bonferroni optimal alpha & 82.7 & 82.7 & 36.5 & 60.0 & 59.2 &   &  &  \\
Bonferroni optimal power (all) & 82.7 & 82.7 & 36.5 & 60.0 & 59.2 &   &  &  \\
Bonferroni greedy algorithm & 82.7 & 82.7 & 36.5 & 59.8 & 59.3 &  &  &  \\
minP & 81.5 & 81.5 & 35.9 & 58.8 & 58.6 &  &  &  &  &  \\
Optimal alpha & 92.6 & 82.3 & 36.5 & 59.5 & 59.3 & 17 & 137 & 889 \\
Optimal area & 93.0 & 84.3 & 36.5 & 60.4 & 60.4 &  17 & 136 & 874 \\
Optimal power (all, $\rho=0$) & 95.7 & 83.9 & 36.5 & 60.2 & 60.1  & 37 & 649 & 4438 \\
Optimal power (all, $\rho=0.5$) & 95.7 & 83.8 & 36.5 & 60.2 & 60.1 &  8 & 43 & 143 \\
Optimal power (EP 1, $\rho=0$) & 92.2 & 82.4 & 36.5 & 60.4 & 58.5 & 43 & 1765 & 14350 \\
Cons. opt. alpha & 84.3 & 84.3 & 36.5 & 60.4 & 60.4 &  0 & 7 & 68 \\
Cons. opt. area & 84.3 & 84.3 & 36.5 & 60.4 & 60.4  & 0 & 7 & 68 \\
Cons. opt. power (all, $\rho=0$) & 84.3 & 84.3 & 36.5 & 60.4 & 60.4  & 0 & 7 & 226 \\
Cons. opt. power (all, $\rho=0.5$) & 84.3 & 84.3 & 36.5 & 60.4 & 60.4  & 0 & 3 & 45 \\
Cons. opt. power (EP 1, $\rho=0$) & 84.3 & 84.3 & 36.5 & 60.4 & 60.4  & 0 & 7 & 242 \\
Greedy algorithm & 93.2 & 84.3 & 36.5 & 60.4 & 60.4 &  &   &  \\
    \bottomrule
    \end{tabular}
  \label{tab:nEP2.n15.alpha0.025.corr0.effall.text}
\end{table}
\clearpage
\end{landscape}

\clearpage
\begin{landscape}
\begin{table}[htbp]
\centering
\caption{Power with three binary endpoints with different effect sizes. The scenario for this table is $n=10$, $\alpha=0.025$, $p_{1,Trt}=0.8,\ p_{2,Trt}=0.7,\ p_{3,Trt}=0.6,\ p_{1,Ctr}=0.2,\ p_{2,Ctr}=0.2,\ p_{3,Ctr}=0.2,\ \rho=0.5$. The caclulations are based on 2000 random samples, the maximal number of iterations in the 
branch and bound algorithm was $5\cdot 10^5$. The table shows the probabilities in percent that the closed testing procedure rejects the global intersection hypothesis $\cap_{i=1}^3 H_i$, at least one elementary null hypothesis (Any $H_i$), all elementary hypotheses simultaneously (All $H_i$), or particularly $H_1$, $H_2$, or $H_3$. The proportion of simulated instances in which a confirmed optimal solution was found by the branch and bound algorithm for all four intersection hypothesis tests in the closed testing procedure is given in the column labelled C(\%). To further quantify the number of required iterations, the median (q50), the 90 \% quantile (q90) and the maximum of the number of iterations is shown in the last three columns. A value of $5 \cdot 10^5$ indicates that the respective quantity to achieve an optimal solution would be above $5 \cdot 10^5$, but is here bounded by the maximal number of iterations.
}
\begin{tabular}{lcccccccccc}
    \toprule
    \multicolumn{1}{c}{Test} & $\cap_{i=1}^3 H_i$ & Any $H_i$ & All $H_i$ & $H_1$ & $H_2$ & $H_3$ & C(\%) & q50 & q90 & Max \\
    \midrule
Bonferroni & 52.8 & 52.8 & 16.1 & 48.7 & 33.5 & 23.0 &  &  &  &  \\
HKT & 55.9 & 55.9 & 16.1 & 51.3 & 34.6 & 23.2 &  &  &  &  \\
Bonferroni optimal alpha & 72.5 & 72.5 & 16.5 & 63.0 & 40.8 & 27.4 &  &  &  &  \\
Bonferroni optimal power & 73.4 & 73.4 & 16.5 & 64.0 & 41.2 & 27.1 &  &  &  &  \\
Bonferroni greedy algorithm & 72.3 & 72.3 & 16.5 & 63.4 & 40.7 & 27.1 &  &  &  &  \\
minP & 72.5 & 72.5 & 16.4 & 63.2 & 40.5 & 27.2 &  &  &  &  \\
Optimal alpha & 78.6 & 66.0 & 16.5 & 57.5 & 38.6 & 27.0 & 99.5 & 44 & 2717.6 & 500000 \\
Optimal area & 80.6 & 73.5 & 16.5 & 64.0 & 41.2 & 27.3 & 100 & 36 & 595.2 & 133909 \\
Optimal power & 84.5 & 72.2 & 16.5 & 63.4 & 40.1 & 27.0 & 100 & 12 & 144 & 24591 \\
Greedy algorithm & 79.2 & 73.8 & 16.5 & 64.2 & 41.2 & 27.5 &  &  &  &  \\
    \bottomrule
    \end{tabular}
\label{tab:power3EP.text}
\end{table}
\end{landscape}
\clearpage

\section{Discussion}
The analysis of small clincial trials is often challenging \cite{EMArare}. Asymptotic methods may lack type I error rate control when the sample size is small, or when there are few events in the case of binary endpoints. Exact tests guarantee type I error rate control, but they are typically overly conservative due to discreteness.
At the same time it is required to make best use of the available information, if the overall number of observations is limited, which favors the analysis of multiple endpoints. This was a motivation for the investigation of optimal rejection regions for multivariate exact tests described in this work. 

Most multiple testing adjustments lead to multivariate rejection regions of a certain restricted shape, e.g. the complement of a hypercube in the case of Bonferroni tests or the minP test. Within a class of shapes, optimization can be performed, however any such shape-restriction leads to increased conservatism when applied to discrete distributions, and as a consequence it has the potential to reduce the power of the test. This limitation can be avoided by allowing for arbitrarily shaped rejection regions. Still, some constraints are required to allow for unambiguous interpretation of the results. In contrast to earlier suggestions for optimal tests with discrete statistics \cite{paroush1969integer,Gutman20072380}, we require that the test decision is monotonic in the value of the test statistic, leading to an additional constraint in the optimization framework.
This condition rules out testing procedures where a (relatively) small observed effect results in rejecting a null hypothesis while a larger effect does not. 
Further, the simulation results of this work compared to those in \cite{Gutman20072380} suggest that the monotonicity constraint (or some similar constraint) is required to obtain a powerful closed testing procedure based on optimal tests for intersection hypotheses.

To control the FWER, it is important to pre-specify the rejection region. Especially, the rejection region must be defined before information on the treatment effect estimates is revealed. In a blinded experiment this means to define the rejection region before unblinding the treatment allocation. In addition, all steps of the  procedure that define the optimized rejection region should be specified in the study protocol.

The choice of the optimization objective function may be based on assumptions about the effect sizes under the alternative. Rejection regions optimized for the true alternative can result in a far more powerful test than those resulting from other optimization criteria. However, as the true alternative is unknown, optimizing power under some assumed alternative is sensitive to having guessed wrongly.
Still, the optimal power test can serve as a useful benchmark to judge the performance of other optimal tests. 
Furthermore, a prior distribution on the effect sizes can be specified and the power averaged over this prior distribution can be optimized.

The reduced power of discrete tests is often attributed to the conservativeness of the test decision under the null hypothesis. Therefore, an obvious choice would be to maximize exhaustion of the type I error rate. This approach indeed results in type I error rates close to the nominal one, but it does not necessarily result in high power. This may be due to the fact that optimizing type I error rate exhaustion does not necessarily lead to rejection regions with a high probability under the alternative.

A useful alternative to the tests derived with the branch and bound algorithm are tests based on the greedy algorithm for multivariate rejection regions with the argmin operator proposed in Section \ref{sec:greedy}. In the scenarios included in the numeric power calculations, this test is often close to optimal. 

Of note, the assumption of exchangeability under the null hypothesis, which was discussed in Section \ref{sec:optfisher}, is required to guarantee FWER control for the procedures relying on the joint conditional distribution of the test statistics. If this assumption is not satisfied, the optimally weighted Bonferroni tests should be preferred over tests using the joint permutation distribution. Among these Bonferroni-type tests, the test based on the greedy algorithm showed robust performance, good power, is $\alpha$-consistent and provides consonant procedures by construction and can therefore be recommended as a good general choice.

While the focus of this paper is on multiple binary endpoints, the proposed theoretical framework for optimal exact tests is more general. It can be applied to multiple hypothesis tests whenever the exact joint distribution of the involved test statistics is known, or, in case of the weighted Bonferroni tests, when the marginal distributions are known. Consider, e.g., the comparison of $k$ treatment groups to a common control with respect to a continuous endpoint, with the aim of showing superiority for at least one treatment versus control. Rank-sum tests may be used as exact marginal tests and optimal rejection regions may be defined for the joint permutation distribution of the $k$ rank-sum statistics. As further example consider testing a treatment effect in $k$ disjoint populations, using $k$ exact tests. The joint distribution of the test statistics then follows from the known marginal distribution and from independence between observations from different populations, and a closed test with optimal local rejection regions can be derived. Similarly, in an analysis involving a full population and a sub-population, the distribution of exact test statistics is given by the known marginal distributions and the correlation structure determined by the proportion of subjects belonging to the sub-population.

In summary, optimizing the rejection region for multivariate exact tests offers a notable advantage over simpler methods in terms of the power to reject a global intersection null hypothesis and, to a lesser extent, the power to reject some elementary null hypothesis. In the small sample setting, where this approach can have the greatest impact, numeric solutions of the discrete optimization problems are found within short computation times. Application of the optimal exact tests may require an additional effort at the planning stage, which is worthwhile if the aim is to make best use of multiple exact hypotheses tests from a small data sample.

\section*{Appendix A - Pre-processing} \label{app:preprocess}

In the first pre-processing step, condition $(i)$ is used to remove all points from the search space $V$ that would inevitably lead to a rejection region with a level greater than $\alpha$. If a point $\boldsymbol{t}=(t_1,\hdots ,t_k ) \in V$ was selected to be part of the rejection region, condition $(ii)$ implies that the set $\{(s_1,\hdots ,s_k ) \in V: s_1 \geq t_1,\hdots ,s_k \geq t_k \}$ is part of the rejection region. Thus the type I error rate of a rejection region containing $\boldsymbol{t}$ is at least $P(\{(s_1,\hdots ,s_k ) \in V: s_1 \geq t_1,\hdots ,s_k \geq t_k \})$. Therefore all points $\boldsymbol{t} \in V$ for which $P(\{(s_1,\hdots ,s_k ) \in V: s_1 \geq t_1,\hdots ,s_k \geq t_k \})>\alpha$ are removed from the search space. Denote the set of the remaining points by $V^{(1)}$.

In the second pre-processing step we identify points that are definitely contained in an optimal level $\alpha$ rejection region, regardless of the optimality criterion. For each point $\boldsymbol{t} \in V^{(1)}$, we calculate an upper bound for the probability mass under $H_0$ of all possible rejection regions that do not contain $\boldsymbol{t}$. If $\boldsymbol{t} \notin R$, the set $A(\boldsymbol{t})=\{(s_1,\hdots ,s_k ) \in V^{(1)}:s_1 \leq t_1,\hdots ,s_k \leq t_k \} \notin R$, because otherwise condition $(ii)$ would be violated. The upper bound for the level when point $\boldsymbol{t}$ is not included in the rejection region is $\alpha_{max,-t}=P_{H_0}(V\backslash A(\boldsymbol{t}))$. If $\alpha_{max,-t} < \alpha$, even the largest rejection region not containing $\boldsymbol{t}$ could possibly be made larger, and the only way to do so is adding $\boldsymbol{t}$, because of condition $(ii)$. So if $\alpha_{max,-t}+P_{H_0}(\boldsymbol{t}) \leq \alpha$, $\boldsymbol{t}$ must be included in each optimal rejection region. Denote the set of the points still remaining after this step by $V^{(2)}$. It is then sufficient to perform the optimization on the remaining search space $V^{(2)}$ for a significance level of $\alpha - P_{H_0}(V^{(1)}\backslash V^{(2)})$.

\begin{figure}[h]
\centering
\begin{center}
      \includegraphics[scale=0.25]{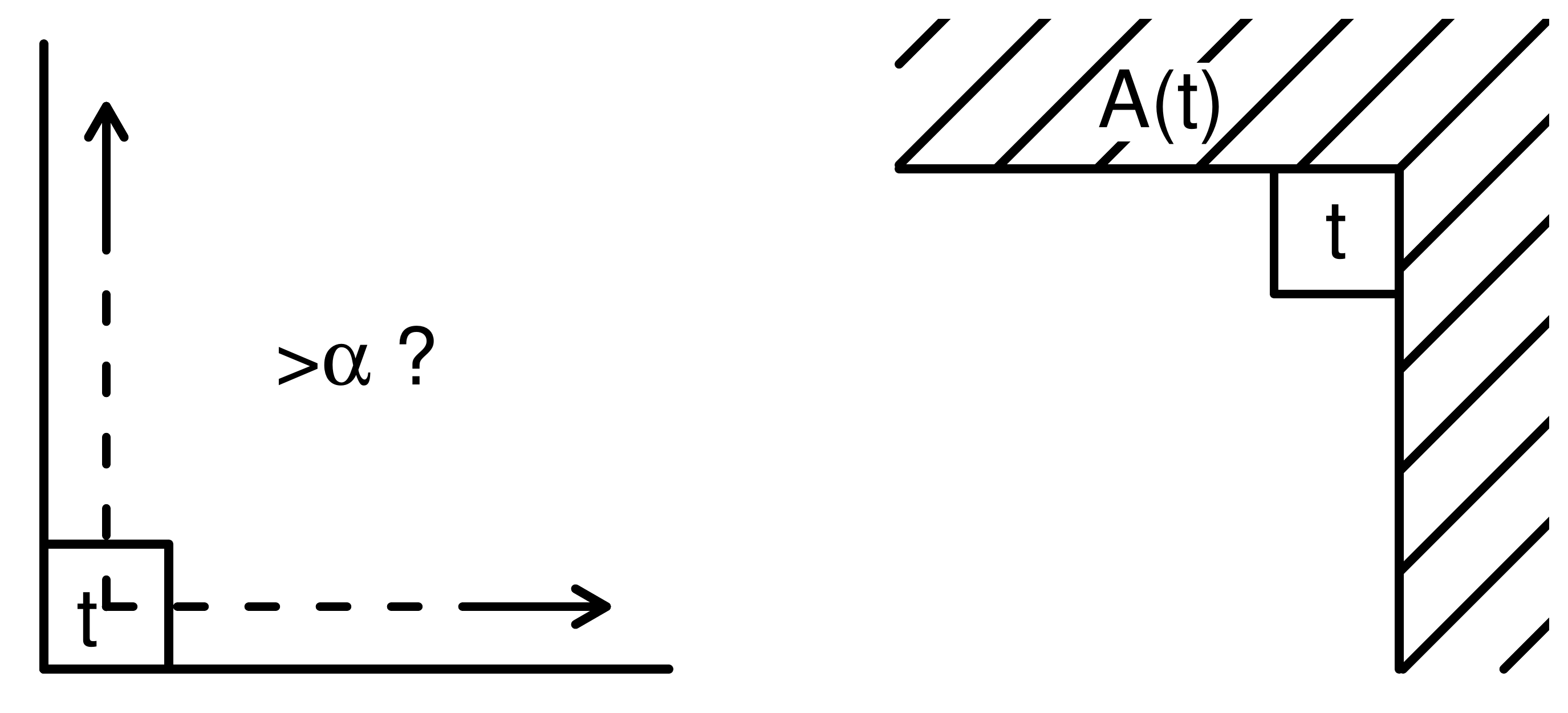}
    \end{center}
   \caption{Left panel: Exclude points $\boldsymbol{t} \in V$ for which $P(\{(s_1,\hdots ,s_k ) \in V: s_1 \geq t_1,\hdots ,s_k \geq t_k \})>\alpha$ from the search space as including these points will result in a level $>\alpha$. 
Right panel: For all points $\boldsymbol{t} \in V$ get the largest region $A(\boldsymbol{t})$ which satisfies condition $(ii)$ and does not include $\boldsymbol{t}$. If $P_{H_0}(A(\boldsymbol{t}))+P_{H_0}(\boldsymbol{t})\leq \alpha$, $\boldsymbol{t}$ must be contained in any optimal rejection region.
  }  \label{Fig:preprocess}
\end{figure}

The two pre-processing steps are illustrated in Figure \ref{Fig:preprocess}.

\section*{Appendix B - Formulation of the optimization problem as integer linear program} \label{app:ILP}
The constraint on the type I error rate $(i)$ can be written as $\boldsymbol{a}^T \boldsymbol{x} \leq \alpha$, where $\boldsymbol{a}=(P_{H_0}(\boldsymbol{t}_1),\ldots,P_{H_0}(\boldsymbol{t}_m))^T$. 
The constraint due to condition $(ii)$ can be written as $C\boldsymbol{x} \leq 0$. Here, $C$ is a matrix with $m$ columns, which captures the requirement that if $x_i=1$, $x_j$ must be one, when $t_{j,l} \geq t_{i,l}$ for all $l=1,\ldots,m$. 
An efficient way to write this constraint matrix requires at most $m$ rows.
Denote by $M_i$ the index set of points that are more extreme than $\boldsymbol{t}_i$, i.e., $M_{i}=\{j=1,\ldots,m:\ t_{j,l} \geq t_{i,l} \text{ for all }l=1,\ldots,m\}$. 
Then we only need one row $\boldsymbol{c}$ with $c_i=|M_{i}|$, $c_j=-1$ for all $j \in M_i$ and all other entries of value 0. Thus $\boldsymbol{c}^T \boldsymbol{x}>0$ if and only if not all points more extreme than $\boldsymbol{t}_i$ are part of the solution, which violates condition $(ii)$. 

Rows corresponding to points for which there are no more extreme points, i.e. $|M_i|=0$, can be removed from the constraint matrix.

Both conditions, $\boldsymbol{a}^T \boldsymbol{x} \leq \alpha$ and $C\boldsymbol{x} \leq 0$ are combined using a final constraint matrix $B=(\boldsymbol{a},C^T)^T$ and a vector $\boldsymbol{b}=(\alpha,0,\ldots,0)^T$, such that the linear optimization problem reads $\boldsymbol{w}^T \boldsymbol{x} \rightarrow \mbox{max, subject to } B\boldsymbol{x} \leq \boldsymbol{b}, \boldsymbol{x} \in \{0,1\}^m$. 

\section*{Appendix C - Separate treatment of points with small probabilities under $H_0$ to reduce computation time} \label{app:small.alpha}
For large optimization problems of the type considered here, a nearly optimal solution can be found with reduced computational effort if points in the search space that have a very small contribution to the type I error rate are treated separately. First, a threshold $c$ considerably smaller than the nominal level of significance is set, e.g. $c=10^{-4}$.
Following the pre-processing, the set of points $C \subseteq V^{(2)}$, such that $max_{\boldsymbol{t} \in C} P_{H_0}(\boldsymbol{t})< min_{\boldsymbol{t} \in V^{(2)} \backslash C} P_{H_0}(\boldsymbol{t})$ and $P_{H_0}(C) \leq c$ is identified. The set $C$ is removed from the search space for the optimization, and the  subsequent optimization is performed on $V^{(2)} \backslash C$ for a significance level of $\alpha - P_{H_0}(V^{(1)}\backslash V^{(2)}) - P_{H_0}(C)$. After the rejection region in $V^{(2)} \backslash C$ has been found, all points in $C$ can be added subject to condition $(ii)$. 

\section*{Appendix D - Binary linear program for Bonferroni-type tests} \label{app:ILP.Bonf}
For appropriate objective functions, linear integer programming may be used to identify the optimal critical boundaries $c_i,i=1,\hdots ,k$. Only values in $V_i^\alpha=\{c \in V_i: S_i(c)\leq \alpha\}$ need to be considered.
As search space for the critical boundaries consider a stacked vector $\boldsymbol{v}=(\boldsymbol{v}_1^T,\hdots ,\boldsymbol{v}_k^T)^T$, with $\boldsymbol{v}_i$ defined as the vector of elements of $V_i^\alpha \cup \infty$.
The solution vector is $\boldsymbol{x} \in \{0,1\}^{|\boldsymbol{v}|}$. The constraint
\begin{equation} \label{marg.constr.1}
	\begin{pmatrix}
		J_1 & 0 & \hdots & 0 \\
		0 & J_2 & \hdots & 0 \\		
		\vdots & \vdots & \vdots & \vdots \\		
		0 & \hdots & 0 &  J_k
	\end{pmatrix}x=
	\begin{pmatrix}
		1 \\ \vdots \\ \vdots \\1
	\end{pmatrix}, \mbox{ with } J_i=(1,\hdots,1)_{1 \times |\boldsymbol{v}_i|}
\end{equation}
ensures that $\boldsymbol{x}$ contains exactly one entry equal to 1 for each search vector $\boldsymbol{v}_i$, indicating the chosen $c_i$. Further, let $\boldsymbol{a}_i$ be the vector of contributions to the type I error rate for the $i$-th endpoint with elements $a_{i,j}=S_i(v_{i,j})$. Let $\boldsymbol{a}=(\boldsymbol{a}_1^T,\hdots ,\boldsymbol{a}_k^T)^T$. Then the constraint
\begin{equation} \label{marg.constr.2}
\boldsymbol{a}^T\boldsymbol{x} \leq \alpha
\end{equation}
guarantees type I error control by the Bonferroni inequality.

The contributions of possible choices for $c_i,i=1,\hdots,k$ to the objective function are formalized similarly in terms of a stacked vector $\boldsymbol{w}=(\boldsymbol{w}_1^T,\hdots,\boldsymbol{w}^T)^T$. Here $w_{i,j}$ is the contribution of the $i$-th test to the objective function if $c_i=v^{(i)}_j$ is selected. Then the objective function is of the type $g=\boldsymbol{w}^T \boldsymbol{x}$, with $\boldsymbol{w}=\boldsymbol{a}$ for objective function (\ref{eq:marg.alpha})
and $w_{i,j}=P_{H_i^{(1)}}(T_i \geq v_{i,j})$ for objective function (\ref{eq:marg.power}).

Thus, the linear integer program is constituted by  $g=\boldsymbol{w}^T \boldsymbol{x}$, the constraints (\ref{marg.constr.1}) and (\ref{marg.constr.2}) and the further constraint that the elements of $x$ are in $\{0,1\}$.

\section*{Appendix E - The conditional distribution of $\YY_{Trt}$}

\[P(\YY_{Trt}=\yy_{Trt}|\YY_{Trt}+\YY_{Ctr}=\mmt) = \frac{P(\YY_{Trt}=\yy_{Trt},\YY_{Trt}+\YY_{Ctr}=\mmt)}{P(\YY_{Trt}+\YY_{Ctr}=\mmt)}=\]
\[=\frac{P(\YY_{Trt}=\yy_{Trt},\YY_{Ctr}=\mmt-\yy_{Trt})}{P(\YY_{Trt}+\YY_{Ctr}=\mmt)}=\frac{P(\YY_{Trt}=\yy_{Trt})P(\YY_{Ctr}=\mmt-\yy_{Trt})}{P(\YY_{Trt}+\YY_{Ctr}=\mmt)}=\]
\[= \frac{1}{P(\YY_{Trt}+\YY_{Ctr}=\mmt)} n_{Trt}! \prod_{s=1}^d \frac{ \left( q_{Trt,s} \right)^{y_{Trt,s}}}{y_{Trt,s}!}
n_{Ctr}! \prod_{s=1}^d \frac{\left( q_{Ctr,s} \right)^{\mt_s-y_{Trt,s}} }{(\mt_s-y_{Trt,s})!} = \]
\[
= \frac{1}{N} \prod_{s=1}^d \frac{1}{y_{Trt,s}!(\mt_s-y_{Trt,s})!}\left( \frac{q_{Trt,s}}{q_{Ctr,s}} \right)^{y_{Trt,s}}
\]

\section*{Acknowledgements}
This paper is based on the Master's thesis of Robin Ristl at the University of Vienna.
We wish to thank Caridad Pontes for pointing us to the clinical trial in patent ductus arteriosus. This work has been funded by the FP7-HEALTH-2013-INNOVATION-1 project Advances in Small Trials Design for Regulatory Innovation and Excellence (ASTERIX) Grant Agreement No. 603160.

\bibliographystyle{unsrtnatmod2}
\bibliography{Literature_multiple_exact_tests_2016_pharma}

\end{document}